\documentclass[conference,nofonttune]{IEEEtran}
\IEEEoverridecommandlockouts

\usepackage{amsthm} 
\usepackage{cite}
\usepackage[pdftex]{graphicx}
\graphicspath{{./figures/}}
\DeclareGraphicsExtensions{.eps}
\usepackage[caption=false,font=footnotesize,labelformat=simple]{subfig}
\usepackage{amsmath}
\usepackage{bm}
\usepackage{color} 
\usepackage{enumitem}
\usepackage{amsfonts}
\usepackage{amssymb,mathrsfs}
\usepackage{array}
\usepackage{latexsym}
\usepackage{microtype} 
\usepackage{cases}
\usepackage{balance}

\usepackage{algorithm}
\usepackage{algpseudocode}
\usepackage{url}

\usepackage{pifont}
\begin{document}

\title{Composite Learning Control With Modular Backstepping and High-Order Tuners
\thanks{*Corresponding Author: Yongping Pan. }
}


\author{
    \IEEEauthorblockN{Tian Shi$^{1}$, Shihua Li$^{1}$, Changyun Wen$^{2}$, and Yongping~Pan$^{*1,2}$}
    \IEEEauthorblockA{$^1$School of Automation, Southeast University, Nanjing 210096, China}
    \IEEEauthorblockA{E-mails: \{shitian,lsh,panyp\}@seu.edu.cn}
    \IEEEauthorblockA{$^2$School of Electrical and Electronic Engineering, Nanyang Technological University, Singapore 639798}
    \IEEEauthorblockA{E-mail: ecywen@ntu.edu.sg}
}

\maketitle

\begin{abstract}
Adaptive control of strict-feedback nonlinear systems with mismatched uncertainties remains challenging because high-order derivatives of parameter estimates can degrade transient tracking, while parameter convergence typically requires the stringent condition of persistent excitation (PE). 
This paper proposes a composite learning backstepping control (CLBC) strategy based on modular backstepping and high-order tuners to achieve closed-loop exponential stability without high-gain feedback and PE. 
A novel composite learning mechanism that maximizes the staged excitation strength is designed for parameter estimation, enabling exponential convergence of the full set of unknown parameters under interval excitation (IE) and exponential convergence of the excited parameter components under partial IE, both of which are strictly weaker than PE. 
An extra prediction error is employed in the adaptive law to ensure transient performance without high-gain feedback. 
Simulations have demonstrated the effectiveness and superiority of the proposed CLBC in both parameter estimation and control compared to state-of-the-art methods.
\end{abstract}

\begin{IEEEkeywords}
Adaptive control, composite learning, exponential stability, mismatched uncertainty, parameter convergence.
\end{IEEEkeywords}

Adaptive control is desirable owing to its unique capacity to accommodate uncertain and time-varying properties of nonlinear systems, and recent survey papers can be found in \cite{Tao2014, Chan2014, Nageshrao2016, Ortega2020, Annaswamy2021, Guo2023, Pan2024LCSS}. 
The presence of mismatched uncertainties is a major obstacle to adaptive control of nonlinear systems. 
Adaptive integral backstepping with overparameterization, which combines integral backstepping and direct adaptation, is a precursor to relaxing the above obstacle \cite{Kanellakopoulos1991}.
A tuning function approach is a direct adaptive backstepping approach without overparameterization \cite[Ch. 4]{Krstic1995}, where 
a tuning function is constructed iteratively at each backstepping step, and an actual adaptive law is generated at the last step using all previous tuning functions. 
The above adaptive backstepping approaches have two major drawbacks: 1) The problem of ``explosion of complexity'' is induced due to the repeated differentiation of virtual control inputs; 2) the exponential stability of the closed-loop system 
relies on a strict condition termed persistent excitation (PE) \cite{Kurdila1995}. 


A modular backstepping approach follows the certainty equivalence principle separating control and estimation designs \cite[Ch. 5]{Krstic1995}. A key feature of this approach is that the time derivatives of virtual control inputs are replaced by their partial derivatives with respect to system states and reference signals, while the resulting high-order time derivatives of parameter estimates are treated as additive disturbances. Therefore, the modular backstepping approach does not involve tuning functions or overparameterization and has lower complexity. 
This approach ensures closed-loop stability and strong robustness by introducing a nonlinear damping term into the stabilizing function at each backstepping step \cite{Wang2011}. A standard gradient-descent identifier derived from a swapping scheme can be combined with modular backstepping to achieve asymptotic tracking \cite[Ch. 6]{Krstic1995}. 
Nevertheless, because the high-order time derivatives of parameter estimates arise in the closed-loop system, the modular backstepping approach may degrade transient tracking and prevent exact parameter estimation, even under the PE condition.

A high-order tuner (HOT) approach can efficiently mitigate the negative effect of the time derivatives of parameter estimates in modular backstepping \cite{Morse1992}.
The key idea of the HOT is to apply a linear filter with a sufficiently high relative degree to the adaptive law, making the exact implementation of high-order time derivatives of parameter estimates feasible.
In \cite{Nikiforov2001}, a direct adaptive control scheme was combined with the HOT to counteract the transient behavior of parameter estimates arising from their high-order time derivatives, thereby improving transient and steady-state tracking performance.
In \cite{Nikiforov2022}, a memory regressor extension (MRE)-based identifier was combined with the HOT to design an indirect adaptive control law, where the HOT is employed to an extended regression model such that the high-order time derivatives of parameter estimates can be calculated exactly without filtering delay. However, the two methods above still require nonlinear damping terms to ensure closed-loop stability and transient performance, and they rely on the stringent PE condition to guarantee exponential stability. 

From the above discussions, existing modular backstepping methods have the following limitations:
\begin{enumerate}
    \item The transient and steady-state tracking performance rely on nonlinear damping terms;
    \item The transient process of parameter estimates due to their high-order time derivatives can destroy the tracking performance and parameter convergence;
    \item Stringent excitation conditions must be satisfied to ensure exponential stability of the closed-loop system.
\end{enumerate}
Motivated by the above facts, this paper proposes a composite learning backstepping control (CLBC) strategy that ensures the exponential stability of the closed-loop system under relaxed excitation conditions for strict-feedback uncertain nonlinear systems. The design procedure is as follows: 
First, the modular backstepping scheme without nonlinear damping is introduced to facilitate the control design; 
second, a generalized regression equation is constructed by the swapping technique with interval integrations; 
third, a linear filter is applied to the generalized regression equation to generate a linearly parameterized model; 
fourth, a generalized prediction error is designed to exploit online data memory; 
fifth, a general prediction error is introduced to counteract a modeling error term;
finally, a composite learning HOT is constructed by combining the two prediction errors to implement the high-order time derivatives of parameter estimates exactly. 
The contributions of this study are twofold:
\begin{enumerate}
\item A feasible modular backstepping strategy termed CLBC is proposed to guarantee transient and steady-state tracking without nonlinear damping terms or high control gains;
\item An algorithm for staged excitation strength maximization is designed to enhance the online data memory of composite learning across different partial excitation stages.
\end{enumerate}

\textit{Notation}: 
$\min\{\cdot\}$ denotes the minimum operator, 
$\sigma_{\min}(A)$ is the minimum singular value of $A$,
$\|\bm x\|$ is the Euclidean norm of $\bm x$, 
$L_\infty$ is the space of bounded signals, 
$I$ is an identity matrix, 
$\bm 0$ is a zero matrix, 
$\Omega_c$ $:=$ $\{\bm x| \|\bm x\| \leq c\}$ is the ball of radius $c$,
$\arg\max_{x\in S} f(x)$ $:=$ $\{x\in S| f(y)$ $\leq$ $f(x)$, $\forall y$ $\in$ $S\}$, 
$\bm g\in\mathcal{C}^k$ indicates that $\bm g$ has continuous partial derivatives up to the order $k$, 
where $A \in \mathbb R^{n \times n}$, $\bm x \in \mathbb R^n$, $c \in \mathbb R^+$, $f: \mathbb R \mapsto \mathbb R$, 
$\bm g$ $:$ $\mathbb{R}^n$ $\mapsto$ $\mathbb{R}^m$,
$S \subset \mathbb R$, $n, m \in \mathbb{N}^+$, and $k\in \mathbb{N}$. 

\section{Problem Formulation}\label{Problem}

Consider a class of $n$th-order strict-feedback uncertain nonlinear systems as follows \cite{Krstic1995}:
\begin{align}\label{eq01}
\left\{\begin{array}{l}
    \dot{x}_i = {\bm\varphi}_i^T({\bm x}_i)\bm\theta + x_{i+1} ,\\
    \dot{x}_n = {\bm\varphi}_n^T(\bm x)\bm\theta + \beta(\bm x)u,\\
    y = x_1
    \end{array}\right.
\end{align}
with $i=$ 1 to $n-1$ and ${\bm x}_i(t) := [x_1(t),x_2(t),\cdots,x_i(t)]^T\in\mathbb{R}^i$, 
where $\bm x(t) :=[x_1(t),x_2(t),\cdots,x_n(t)]^T$ $\in\mathbb{R}^n$ is a system state, $u(t)\in\mathbb{R}$ is a control input, $y(t)\in\mathbb{R}$ is a system output, $\bm\theta\in\Omega_{{\rm c}_\theta}\subset\mathbb{R}^N$ with ${\rm c}_\theta\in\mathbb{R}^+$ is an unknown parameter vector, 
${\bm\varphi}_i: \mathbb{R}^i$ $\rightarrow\mathbb{R}^N$ denotes a known regressor, $\beta:$ $\mathbb{R}^n\rightarrow\mathbb{R}$ is a known gain function, and $N$ is the number of parameter elements.
Let $y_{\rm r}(t) \in \mathbb R$ be a reference signal. 
The following definitions are given for the subsequent analysis.

\textit{Definition 1} \cite{Sastry1989a}: A bounded regressor $\Phi(t) \in \mathbb R^{N\times n}$ is of PE if there exist constants $t_0$, $\sigma$, and $\tau_{\rm d} \in \mathbb R^+$ such that
\[\int_{t-\tau_{\rm d}}^{t}\Phi(\tau)\Phi^T(\tau)d\tau \geq \sigma I, \forall t \geq t_0.\]

\textit{Definition 2} \cite{Kreisselmeier1990}: A bounded regressor $\Phi(t) \in \mathbb R^{N\times n}$ is of IE if there exist constants $\sigma$, $\tau_{\rm d}$, and $T_{\rm e} \in \mathbb R^+$ such that
\[\int_{T_{\rm e}-\tau_{\rm d}}^{T_{\rm e}}\Phi(\tau)\Phi^T(\tau)d\tau \geq \sigma I.\]

\textit{Definition 3:} A bounded regressor $\Phi(t) \in \mathbb R^{N\times n}$ is of partial IE, if there exist constants $\sigma$, $\tau_{\rm d}$, and $T_{\rm a}$ $\in$ $\mathbb R^+$ such that 
\[\int_{T_{\rm a}-\tau_{\rm d}}^{T_{\rm a}} \Phi_\zeta(\tau)\Phi_\zeta^T(\tau)d\tau \geq \sigma I\] 
in which $\Phi_\zeta \in \mathbb{R}^{m\times n}$ is a sub-regressor obtained by eliminating some rows of $\Phi$ with $1\leq m < N$.

For convenience, a column ${\bm\phi}_{j}(t)\in\mathbb{R}^n$ ($j=1$ to $N$) of a regressor $\Phi^T(t)\in\mathbb{R}^{n\times N}$ is named as a channel. Consequently, one has $\Phi(t)$ $=$ $[{\bm\phi}_1(t), {\bm\phi}_2(t), \cdots, {\bm\phi}_N(t)]^T$. A channel ${\bm\phi}_j(t)$ is named as an \textit{active channel} if $\|{\bm\phi}_j(t)\| \neq 0$, conversely termed as an \textit{inactive channel}.
Without loss of generality, assume that there exists a proper time window $\tau_{\rm d}$ satisfying either Definition 2 or Definition 3, and consider the case where IE may not exist, $\forall t \geq 0$, but partial IE exists at the beginning and some moments later.
We aim to design a suitable adaptive control strategy for the system \eqref{eq01} that guarantees closed-loop stability and parameter convergence without the PE condition.


\section{Modular Backstepping Control Design}\label{Design} 

The following assumption presented in \cite{Krstic1995} is given for the modular backstepping control design. 

\textit{Assumption 1:} $y_{\rm r}$, $\dot y_{\rm r}$, $\cdots$, $y^{(n)}_{\rm r}$ $\in$ $L_\infty$, $\beta(\bm x)\neq 0$, $\forall \bm x\in\mathbb{R}^n$, and $\bm\varphi_i\in\mathcal{C}^{n-i}$ for $i=1$ to $n$.

The virtual control inputs $v_1(t), v_i(t)\in\mathbb{R}$ in the modular backstepping approach are recursively given by \cite{Krstic1995}
\begin{subequations}\label{eq03}
\begin{align}
    &v_1(x_1,\hat{\bm\theta},y_{\rm r}) = -k_{{\rm c}1}e_1 - \bm\psi_1^T\hat{\bm\theta},\\
&v_{i}({\bm x}_{i},\Theta_{i-1},{\bm y}_{{\rm r}i}) = - k_{{\rm c}i}e_{i} - e_{i-1} - \bm\psi_{i}^T\hat{\bm\theta}\notag\\
    +&\sum_{k=1}^{i-1}\left(\frac{\partial v_{i-1}}{\partial x_k}x_{k+1} + \frac{\partial v_{i-1}}{\partial\hat{\bm\theta}^{(k-1)}}\hat{\bm\theta}^{(k)} + 
    \frac{\partial v_{i-1}}{\partial y_{\rm r}^{(k-1)}}y_{\rm r}^{(k)}\right)
\end{align}
\end{subequations}
with $\Theta_{i-1}(t)$ $:=$ $\big[\hat{\bm\theta}(t), \dot{\hat{\bm\theta}}(t), \cdots, \hat{\bm\theta}^{(i-1)}(t)\big]$ $\in\mathbb{R}^{N\times i}$ and ${\bm y}_{{\rm r}i}(t)$ $:=$ $\big[y_{\rm r}(t)$, $\dot{y}_{\rm r}(t)$, $\cdots$, $y^{(i-1)}_{\rm r}(t) \big]^T$ $\in\mathbb{R}^{i}$, where $e_1(t)$ $:=$ $x_1(t) - y_{\rm r}(t)\in\mathbb{R}$ and $e_{i}(t)$ $:=$ $x_{i}(t) - v_{i-1}(t) - y_{\rm r}^{(i-1)}(t)\in\mathbb{R}$ are tracking errors,
$\bm\psi_1 := \bm\varphi_1$ and
$\bm\psi_i := \bm\varphi_i - \sum_{k=1}^{i-1}\frac{\partial v_{i-1}}{\partial x_k}\bm\varphi_k \in\mathbb{R}^N$
are regressors, $k_{{\rm c}1}, k_{{\rm c}i}\in\mathbb{R}^+$ are control gain parameters, and $i=2, \cdots, n$. In the final step, design the control law
\begin{align}\label{eq05}
    u = \frac{1}{\beta(\bm x)}\left(v_n(\bm x, \Theta_{n-1},{\bm y}_{{\rm r}n}) + y_{\rm r}^{(n)}\right).
\end{align}
Applying \eqref{eq03}, \eqref{eq05}, and $e_{i}$ $=$ $x_{i} - v_{i-1} - y_{\rm r}^{(i-1)}$ to \eqref{eq01} results in a closed-loop tracking error system
\begin{align}\label{eq06}
    \dot{\bm e} = \Lambda\bm e + \Phi^T(\bm x, \Theta_{n-1}, {\bm y}_{{\rm r}n})\tilde{\bm\theta}
\end{align}
in which $\bm e(t)$ $:=$ $[e_1(t),e_2(t),\cdots,e_n(t)]^T\in\mathbb{R}^n$ is a tracking error vector, $\Phi$ $:=$ $[\bm\psi_1$, $\bm\psi_2$, $\cdots$, $\bm\psi_n]$ $\in\mathbb{R}^{N\times n}$ is a new regressor, $\tilde{\bm\theta}(t)$ $:=$ $\bm\theta - \hat{\bm\theta}(t)$ $\in\mathbb{R}^N$ is a parameter estimation error, and
\begin{align*} 
    \Lambda = \left[\begin{matrix}
    -k_{{\rm c}1}     & 1       &  0      &  \cdots  & 0 \\
    -1       & -k_{{\rm c}2}    &  1      &  \cdots  & 0 \\
    \vdots   &  \vdots &  \vdots &  \ddots  & \vdots \\
    0        &  0      &  0      &  \cdots  & 1 \\
    0        &  0      &  0      &  \cdots  & -k_{{\rm c}n} \\
    \end{matrix}\right] \in\mathbb{R}^{n\times n}.
\end{align*}

Consider linear filtering operations
\begin{align}\label{eq08}
    \left\{\begin{array}{ll}
    \dot{\bm\zeta}(t) = \Lambda\bm\zeta(t) + \Phi^T(t)\hat{\bm\theta}(t), &\bm\zeta(0) = -\bm e(0)\\
    \dot{\Phi}_{\rm s}^T(t) =  \Lambda{\Phi}_{\rm s}^T(t) + \Phi^T(t), &{\Phi}_{\rm s}(0) = \bm 0
    \end{array}
    \right.
\end{align}
and define an output vector $\bm p(t)$ $:=$ $\bm e(t) + \bm\zeta(t) \in \mathbb{R}^n$, where $\bm\zeta(t) \in \mathbb{R}^{n}$ and ${\Phi}_{\rm s}(t) \in\mathbb{R}^{N\times n}$ are filtered outputs. Following the swapping technique \cite[Ch. 6]{Krstic1995} and \eqref{eq08}, one can obtain a static linear parametric model as a regression equation as follows:
\begin{align}\label{eq09}
    \bm p(t) = \Phi^T_{\rm s}(t)\bm\theta.
\end{align}
However, the parameter vector $\bm\theta$ in \eqref{eq09} cannot be estimated by classical adaptation schemes because the regressor $\Phi_{\rm s}$ relies on the inaccessible high-order time derivatives $\hat{\bm\theta}^{(k)}$ ($k=1$ to $n-1$).
Even when $\hat{\bm\theta}^{(k)}$ are available, the parameter convergence of classical adaptation schemes relies on the stringent PE condition, which requires that the reference trajectory $y_{\rm r}$ includes sufficiently rich spectral information at all times.

\section{Composite Learning Design}\label{Analysis}


For convenience, let $\Phi_{{\rm s},\zeta}$ $:=$ $[\bm\phi_{{\rm s},k_1}, \bm\phi_{{\rm s},k_2}, \cdots, \bm\phi_{{\rm s},k_{N_\zeta}}]^T$ $\in\mathbb{R}^{N_\zeta\times n}$ ($1\leq k_j\leq N$, $j=1$ to $N_\zeta$) be an active sub-regressor of $\Phi_{\rm s}$ in \eqref{eq09}, $N_\zeta<N$ be the number of active channels, and $\bm\psi_{\zeta, i}$ $\in\mathbb{R}^{N_\zeta}$ ($i=1$ to $n$) be the $i$th column of $\Phi_{{\rm s},\zeta}$. The following partial identifiability assumption is introduced to facilitate the composite learning HOT design and to ensure the existence of partial IE at the beginning and at some later moments.

\textit{Assumption 2:} There exists at least one regressor vector $\bm\psi_{\zeta, i}$ $\in$ $\mathbb{R}^{N_\zeta}$ and a set of time instants $\{t_j\}$ with $t_j\in[T_{\rm a}-\tau_{\rm d}, T_{\rm a}]\subset\mathbb{R}^+$ to get rank$\{\bm\psi_{\zeta, i}(t_1),\bm\psi_{\zeta, i}(t_2),\cdots, \bm\psi_{\zeta, i}(t_{N_\zeta})\}$ $=$ $N_\zeta$.

Multiplying \eqref{eq09} by $\Phi_{\rm s}$ and letting $\bm\zeta(0) = -\bm e(0)$ and ${\Phi}_{\rm s}(0) = \bm 0$, one obtains an extended regression equation 
\begin{align}\label{eq_er}
    \Phi_{\rm s}(t)\Phi_{\rm s}^T(t)\bm\theta = \Phi_{\rm s}(t) \bm p(t).
\end{align}
Integrating \eqref{eq_er} over a moving time window $[t-\tau_{\rm d},t]$, one obtains a generalized regression equation
\begin{align}\label{eq11b}
    \Psi(t)\bm\theta = \bm q(t)
\end{align}
where $\Psi(t) \in \mathbb R^{N\times N}$ is an excitation matrix given by
\begin{align}\label{eq10}
    \Psi(t) := \int_{t-\tau_{\rm d}}^t\Phi_{\rm s}(\tau)\Phi_{\rm s}^T(\tau)d\tau
\end{align}
and $\bm q(t) \in \mathbb R^{N}$ is an auxiliary variable given by
\begin{align}\label{eq12}
    \bm q(t) := \int_{t-\tau_{\rm d}}^t\Phi_{\rm s}(\tau)\bm p(\tau)d\tau.
\end{align}
To obtain the high-order time derivatives $\hat{\bm\theta}^{(k)}$, it is feasible to apply a linear filter with $n-1$ relative degrees
\begin{align}\label{eq15}
    H(s) := \prod_{i=1}^{n-1}\frac{\alpha_i}{s+\alpha_i}
\end{align}
with $\alpha_i\in\mathbb{R}^+$ ($i =$ 1 to $n-1$) being filtering constants to \eqref{eq11b}, which results in a generalized parameterized model
\begin{align}\label{eq11c}
    Q(t)\bm\theta = {\bm q}_{\rm f}(t)
\end{align}
where $Q(t):= H(s)[\Psi(t)]$ and ${\bm q}_{\rm f}(t) := H(s)[\bm q(t)]$. Then, an adaptive law of $\hat{\bm\theta}$ can be designed such that $\hat{\bm\theta}^{(k)}$ is obtainable by the direct differentiation of filtered elements on $\Psi$ and $\bm q$. 

From Assumption 2, partial IE exists at the beginning and some later moments, and there exist some constants $\sigma$, $\tau_{\rm d}$, and $T_{\rm a}$ $\in\mathbb{R}^+$ such that
$\Psi_\zeta(T_{\rm a}) \geq \sigma I$
with
\begin{equation}\label{eq11d}
    \Psi_\zeta(t):= \int_{t-\tau_{\rm d}}^t\Phi_{\rm s,\zeta}(\tau)\Phi_{\rm s,\zeta}^T(\tau)d\tau
\end{equation}
in which $\Phi_{\rm s,\zeta}\in\mathbb{R}^{N_\zeta\times n}$ is a sub-regressor composed of all active channels ${\bm\phi}_{{\rm s},k_j}$ of $\Phi_{\rm s}$, i.e., 
$\Phi_{\rm s,\zeta}:=[{\bm\phi}_{{\rm s},k_1} ,{\bm\phi}_{{\rm s},k_2},\cdots,{\bm\phi}_{{\rm s},k_{N_\zeta}}]^T$ with $\|{\bm\phi}_{{\rm s},k_j}(\tau_j)\|>0$, $\exists\tau_j\in[t-\tau_{\rm d},t]$, $1\leq k_j\leq N$, and $j=1$ to $N_\zeta$. 
If the IE condition holds, then there exists a finite time $T_{\rm e}\in\mathbb{R}^+$ such that $\Psi(T_{\rm e})\geq\sigma I$; otherwise, $T_{\rm e}=\infty$.
Note that in practice, active channels can be determined by 
$\Psi_{k_j,k_j}(t)\geq\mu$ with a small threshold $\mu\in\mathbb{R}^+$ instead of $\|\bm\phi_{{\rm s},k_j}\|>0$, where $\Psi_{k_j,k_j}(t)$ is the $k_j$th diagonal element of $\Psi(t)$ in \eqref{eq10}. This is because measurement noise may prevent $\|\bm\phi_{{\rm s},k_j}\|$ from being exactly zero even for inactive channels. The small threshold $\mu$ is introduced to ensure the practical implementability of adaptive control, whereas the ideal condition $\|\bm\phi_{{\rm s},k_j}\|=0$ is adopted in the theoretical analysis to keep it concise.

The index $k_j$ of the active channels $\bm\phi_{{\rm s},k_j}$ may change under partial IE, leading to multiple partial IE stages. To consider the changes of the sub-regressor $\Phi_{{\rm s},\zeta}$ under different partial IE stages, let $\mathcal{I}$ $:=$ $\{k_1, k_2, \cdots, k_{N_\zeta}\}$ and $\mathcal{I}'$ $:=$ $\{k_1', k_2', \cdots, k_{N_\zeta'}'\}$ ($1\leq k_j'\leq N$ and $j=1$ to $N_\zeta'$) be index sets of active channels in the current and previous partial IE stages, respectively, where $N_\zeta'<N$ is the number of previous active channels. 
Algorithm 1 is provided to reconstruct the sub-regressor $\Phi_{{\rm s},\zeta}$ and maximize the exciting strength $\sigma_{\min}(\Psi_\zeta(t))$ in each partial IE stage, where 
$T_{\rm s} \in\mathbb{R}^+$ is a sampling time, $T_{\rm a}$ is the first epoch in each partial IE stage,
$\sigma_{\rm c}(t) \in\mathbb{R}^+$ is the current maximal exciting strength, and $t_{\rm e} \in\mathbb{R}^+$ is the corresponding exciting time. Based on the above argument, define a generalized prediction error
\begin{align}\label{eq14}
    \bm\xi(t) := {\bm q}_{\rm f}(t, t_{\rm e}) - Q(t, t_{\rm e})\hat{\bm\theta}(t)
\end{align} 
with $Q(t, t_{\rm e})$ $:=$ $H(s)[\Psi(t_{\rm e})]$ and ${\bm q}_{\rm f}(t, t_{\rm e})$ $:=$ $H(s)[{\bm q}(t_{\rm e})]$. 

\begin{figure}[!tbp]
    \centering
    \includegraphics[width=3.45in]{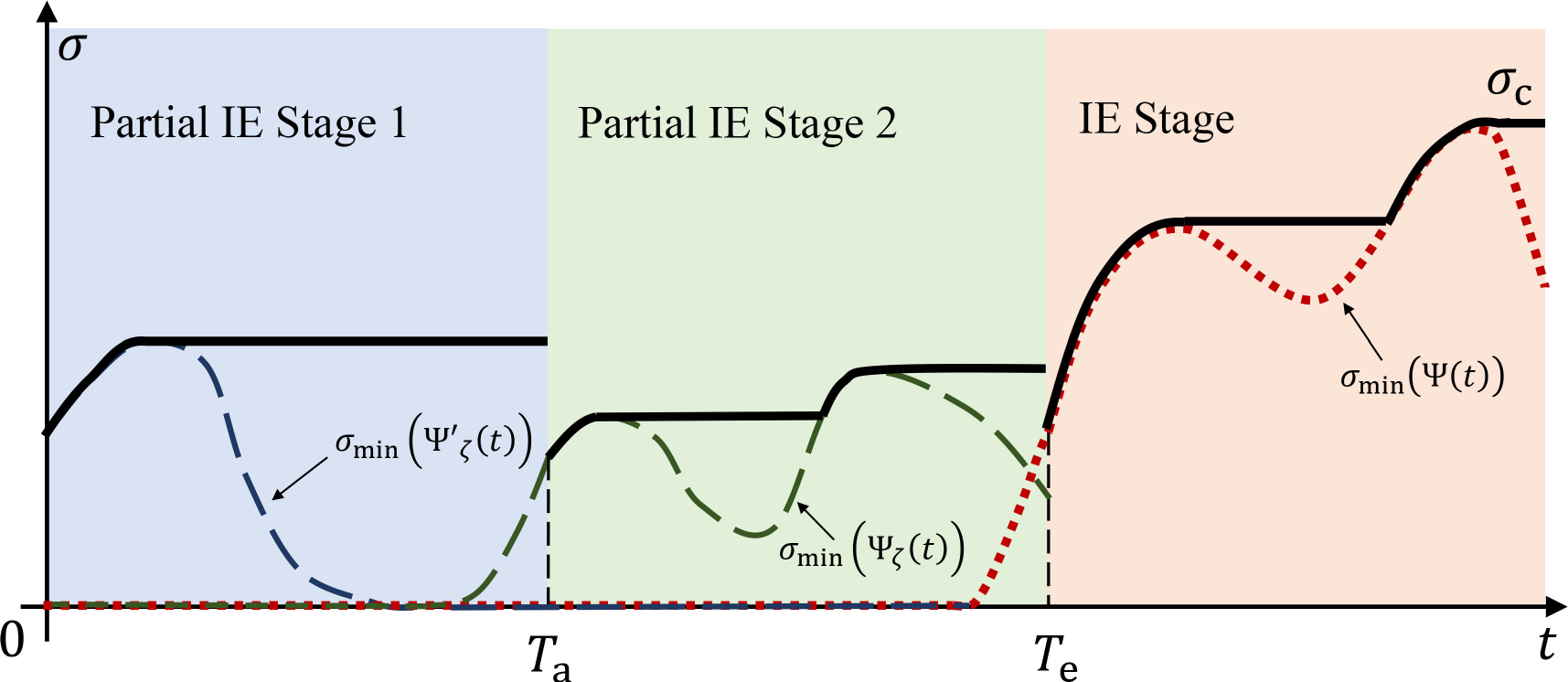}
    \caption{An illustration of the current maximal exciting strength $\sigma_{\rm c}$ in Algorithm 1. Note that the black solid line denotes $\sigma_{\rm c}$, the blue and green dashed lines are $\sigma_{\min}(\Psi_\zeta)$ in two partial IE stages, and the red dotted line is $\sigma_{\min}(\Psi)$.}
    \label{fig:sigma}
\end{figure}

In Algorithm 1, we first choose a sufficiently small threshold $\sigma$ [see Line 1 in Algorithm 1]. 
Then, at the beginning of each partial IE stage, the maximal exciting strength $\sigma_{\rm c}$ is reset to $\sigma$ [see Line 7 in Algorithm 1]. If the current exciting strength $\sigma_{\min}(\Psi_\zeta(t))$ is greater than $\sigma_{\rm c}$, $t_{\rm e}$ and $\sigma_{\rm c}$ are updated [see Lines 9--11 in Algorithm 1]; otherwise, they remain unchanged. Thus, Algorithm 1 ensures that the exciting strength $\sigma_{\min}(\Psi_\zeta(t_{\rm e}))$ is monotonically non-decreasing at each partial IE stage. 
Fig. \ref{fig:sigma} illustrates $\sigma_{\rm c}$ in Algorithm 1 for a simple case with two partial IE stages. As new active channels $\bm\phi_{{\rm s},k_j}(t)$ exist at $t=T_{\rm a}$, the sub-regressor $\Phi_{{\rm s},\zeta}$ with the excitation matrix $\Psi_\zeta(t)$ in \eqref{eq11d} is reconstructed by all new active channels [see Lines 4--8 in Algorithm 1]. 
In this partial IE stage, as the exciting strength $\sigma_{\min}(\Psi_\zeta(t))$ can vary [see the green dashed line in Fig. \ref{fig:sigma}], $t_{\rm e}$ is updated using $\max_{\tau\in[T_{\rm a},t]}\sigma_{\min}(\Psi_\zeta(\tau))$ (exciting strength maximization) [see Lines 9--11 in Algorithm 1 and the black solid line in Fig. \ref{fig:sigma}]. 
This is the same for the IE stage [see Lines 13--15 in Algorithm 1 and the red area in Fig. \ref{fig:sigma}].

\begin{algorithm}[!t]
\caption{Staged exciting strength maximization}
\label{alg:MC-CLRC}
\begin{algorithmic}[1]
\State \textbf{Initialize}: $\mathcal{I}' \leftarrow \emptyset$, $T_{\rm a}=0$, $\sigma_{\rm c} = \sigma$, $t_{\rm e}=0$, $\mu\in\mathbb{R}^+$
 \For{$t = 0$ with a step size of $T_{\rm s}$}
    \If{length$(\mathcal{I}')<N$}
        \State Find the indexes $k_j$ satisfying $\Psi_{k_j,k_j}(t) \leq \mu$, $k_j \in \{1, 2, \cdots, N\}$
        and set $\mathcal{I} \leftarrow \{k_1,k_2,\cdots,k_{N_\zeta}\}$
        \If{$\exists k_j\in\mathcal{I}$ such that $k_j\notin\mathcal{I}'$}
            \State Reconstruct $\Phi_{\rm s,\zeta}(t)$ and $\Psi_\zeta(t)$ by $\mathcal{I}$
            \State $\sigma_{\rm c}\leftarrow\sigma$, $T_{\rm a}\leftarrow t$,
            $\mathcal{I}'\leftarrow\mathcal{I}$
        \EndIf
        \If{$\sigma_{\min}(\Psi_\zeta(t))\geq\sigma_{\rm c}$}
            \State $\sigma_{\rm c}\leftarrow\sigma_{\min}(\Psi_\zeta(t))$, $t_{\rm e}\leftarrow t$
        \EndIf
    \Else 
        \If{$\sigma_{\min}(\Psi(t))\geq\sigma_{\rm c}$}
            \State $\sigma_{\rm c}\leftarrow\sigma_{\min}(\Psi(t))$, $t_{\rm e}\leftarrow t$
        \EndIf
    \EndIf
\EndFor
    \end{algorithmic}
\end{algorithm}

To counteract the transient process of the modeling error term $\Phi^T\tilde{\bm\theta}$ in \eqref{eq06} such that closed-loop stability can be ensured without resorting to the nonlinear damping terms $k_{\rm{d} \it i}\|\bm\psi_i\|^2$ ($i =$ 1 to $n$), we introduce extra prediction error feedback. 
Applying $H(s)$ in \eqref{eq15} to \eqref{eq06} results in a filtered regression equation
\begin{align}\label{eq16}
   \bm z(t) = \Phi_{\rm f}^T(t)\bm\theta 
\end{align}
with $\bm z(t)$ $:=$ $sH(s)[\bm e] - h(t){\bm e}(0) + H(s)[\Phi^T\hat{\bm\theta} - \Lambda\bm e]$ and $\Phi_{\rm f}$ $:=$ $H(s)[\Phi]$, where $h(t):=\mathcal{L}^{-1}\{H(s)\}$ denotes the impulse response of $H(s)$. Then, giving a filtered prediction model
\begin{align}\label{eq16a}
    \hat{\bm z}(t) = \Phi_{\rm f}^T(t)\hat{\bm\theta}(t)
\end{align}
define a general filtered prediction error
\begin{align}\label{eq17}
    \bm\epsilon(t) := \bm z(t) - \Phi_{\rm f}^T(t)\hat{\bm\theta}(t)
\end{align}
where $\hat{\bm z}(t)\in\mathbb{R}^n$ is a predicted value of $\bm z(t)$.
 
With the generalized prediction error $\bm\xi$ in \eqref{eq14} and the filtered prediction error $\bm\epsilon$ in \eqref{eq17}, design a composite learning HOT
\begin{align}\label{eq_lsclbc}
    \dot{\hat{\bm\theta}} = 
    \kappa_1\Phi_{\rm f}\bm\epsilon(t) + \kappa_2\bm\xi(t)
\end{align}
in which 
$\kappa_1$, $\kappa_2\in\mathbb{R}^+$ are weighting factors.  
Since $H(s)$ in \eqref{eq15} has $n-1$ relative degrees, the time derivatives of $\hat{\bm\theta}$ in \eqref{eq_lsclbc} up to the ($n-1$)th order can be implemented physically by a direct differentiation scheme \cite{Gerasimov2020}. More specifically,
the high-order time derivatives of $\hat{\bm\theta}$ are calculated by 
\begin{align}\label{eq20}
    \hat{\bm\theta}^{(k+1)} = 
    \kappa_1\sum_{i=0}^{k}C_k^i s^{k-i}H(s)[\Phi]\bm\epsilon^{(i)} + \kappa_2\bm\xi^{(k)}
\end{align}
with $\hat{\bm\theta}^{(i)}(0)$ $=$ $\bm 0$ and intermediate time derivatives
\begin{align*}
    &\bm\epsilon^{(i)} = \bm z^{(i)} -\sum_{j=0}^{i}C_i^js^{i-j}H(s)[\Phi]\hat{\bm\theta}^{(j)},\\
    &\bm\xi^{(k)} = {\bm q}_{\rm f}^{(k)}(t,t_{\rm e}) - \sum_{i=0}^kC_k^iQ^{(k-i)}(t,t_{\rm e})\hat{\bm\theta}^{(i)}
\end{align*}
in which $C_k^i = k!/(i!(k-i)!)$ are binomial coefficients with $0\leq i\leq k$ and $1\leq k\leq n-2$. 


\section{Theoretical Guarantees}

\subsection{Parameter Convergence Results}

Let $\Phi_{{\rm f},\zeta}$ $:=$ $[\bm\phi_{{\rm f},k_1}, \bm\phi_{{\rm f},k_2}, \cdots, \bm\phi_{{\rm f},k_{N_\zeta}}]^T$ $\in\mathbb{R}^{N_\zeta\times n}$ denote an active sub-regressor of $\Phi_{\rm f}$ in \eqref{eq16a}, in which $\bm\phi_{{\rm f},k_j}$ is the $k_j$th channel. Define a parameter estimation error $\tilde{\bm\theta}_\zeta$ $:=$ $[\tilde{\theta}_{k_1}, \tilde{\theta}_{k_2}, \cdots$, $\tilde{\theta}_{k_{N_\zeta}}]^T$ $\in$ $\mathbb{R}^{N_\zeta}$ regarding active channels. The following theorem presents the parameter convergence results of this study, where proofs are omitted to save space.

\textit{Theorem 1:} Let $[0, t_{\rm f})$ with $t_{\rm f} \in (0,\infty]$ be the maximal time interval on which solutions to the system \eqref{eq01} exist. For any given unknown $\bm\theta\in\Omega_{{\rm c}_\theta}$, the composite learning law \eqref{eq_lsclbc} ensures:
\begin{enumerate}
\item The estimation error $\tilde{\bm\theta}(t)$ is of $L_\infty$, and the prediction error $\bm\epsilon(t)$ in \eqref{eq17} is of $L_2\cap L_\infty$, $\forall t\in [0,t_{\rm f})$;
\item The partial estimation error $\tilde{\bm\theta}_\zeta(t)$ $\rightarrow$ $\bm 0$ exponentially if $t_{\rm f} = \infty$ and partial IE exists for constants $\sigma$, $T_{\rm a}\in\mathbb{R}^+$ and the index set $\mathcal{I}$ no longer changes on $t \in (T_{\rm a}, \infty)$;
\item The estimation error $\tilde{\bm\theta}(t)$ $\rightarrow$ $\bm 0$ exponentially if $t_{\rm f} = \infty$ and IE exists for constants $\sigma$, $T_{\rm e}\in\mathbb{R}^+$.
\end{enumerate}


\subsection{Closed-Loop Stability Results}




The following theorem establishes the stability of the closed-loop system \eqref{eq06} subject to \eqref{eq_lsclbc}.

\textit{Theorem 2:} For the system \eqref{eq01} under Assumptions 1--2 driven by the CLBC law \eqref{eq05} and \eqref{eq_lsclbc} with $\bm x(0) \in\Omega_{{\rm c}_{0}}$ and $\hat{\bm\theta}(0)\in\Omega_{{\rm c}_\theta}$, there exist proper control parameters $k_{{\rm c}1}$ to $k_{{\rm c}n}$ in \eqref{eq03} and filtered parameters $\alpha_1$ to $\alpha_{n-1}$ in \eqref{eq15}, such that 
\begin{enumerate}
\item The closed-loop system \eqref{eq06} with \eqref{eq_lsclbc} is uniform ultimate bounded (UUB) on $t$ $\in$ $[0, \infty)$;
\item The tracking error $\bm e$ and the partial estimation error $\tilde{\bm\theta}_\zeta$ converge to $\bm 0$ exponentially for $t$ $\geq$ $T_{\rm a}$ if partial IE in Definition 3 exists for constants $T_{\rm a}$, $\sigma$ $\in$ $\mathbb{R}^+$ and the index set $\mathcal{I}$ no longer changes on $t \in (T_{\rm a}, \infty)$; 
\item The tracking error $\bm e$ and the estimation error $\tilde{\bm\theta}$ converge to $\bm 0$ exponentially for $t$ $\geq$ $T_{\rm e}$ if IE in Definition 2 exists for constants $T_{\rm e}$, $\sigma\in\mathbb{R}^+$. 
\end{enumerate}

\section{Simulation Studies}\label{Simulation}


This section is devoted to verifying the exponential stability and parameter convergence of the proposed CLBC in \eqref{eq05} with \eqref{eq_lsclbc} under various excitation conditions.
Consider a mass-spring-damping model as follows \cite{Adetola2010}:
\begin{align*}
\left\{\begin{array}{l}
    \dot{x}_1 = x_2, \\
    \dot{x}_2 = x_3 + \bm\varphi_2^T(\bm x_2)\bm\theta,\\
    \dot{x}_3 = u,\\
    y = x_1
\end{array}\right.
\end{align*}
where $x_1\in\mathbb{R}$ denotes a mass position, $u\in\mathbb{R}$ is a control input, and $\bm\theta\in\mathbb{R}^3$ is an unknown parameter vector. Noting \eqref{eq01}, one has
$\bm\varphi_2({\bm x}_2) = [-x_2,-x_1,-x_2^3]^T$ and $\bm\varphi_1(x_1) = \bm\varphi_3({\bm x}) = \bm 0$. 

Set the control parameters $k_{{\rm c}1}=k_{{\rm c}2}=k_{{\rm c}3} =$ 1, $\kappa_1 = \kappa_2$ $=$ $\tau_{\rm d} = 3$, $\hat{\bm\theta}(0) = \bm 0$, and $\sigma=10^{-4}$, and the stable filter $H(s) = 25/(s^2+10s+25)$ in \eqref{eq15}. Gaussian white noise with mean 0 and standard deviation 0.001 is added to the measurements of the system states $x_i$.
The MRE-HOT in \cite{Nikiforov2022} and the composite learning dynamic surface control (CL-DSC) in \cite{Pan2016ijrnc} are selected as baseline controllers, with the damping parameters $k_{\rm{d} \it i} =$ 0.1 for the MRE-HOT, the stable filter $L(s)$ $=$ $20/(s+20)$ for the CL-DSC, and the other shared parameters being the same values as the proposed CLBC for fair comparisons.

\begin{figure}[!t]
\centering
\includegraphics[width=3.4in]{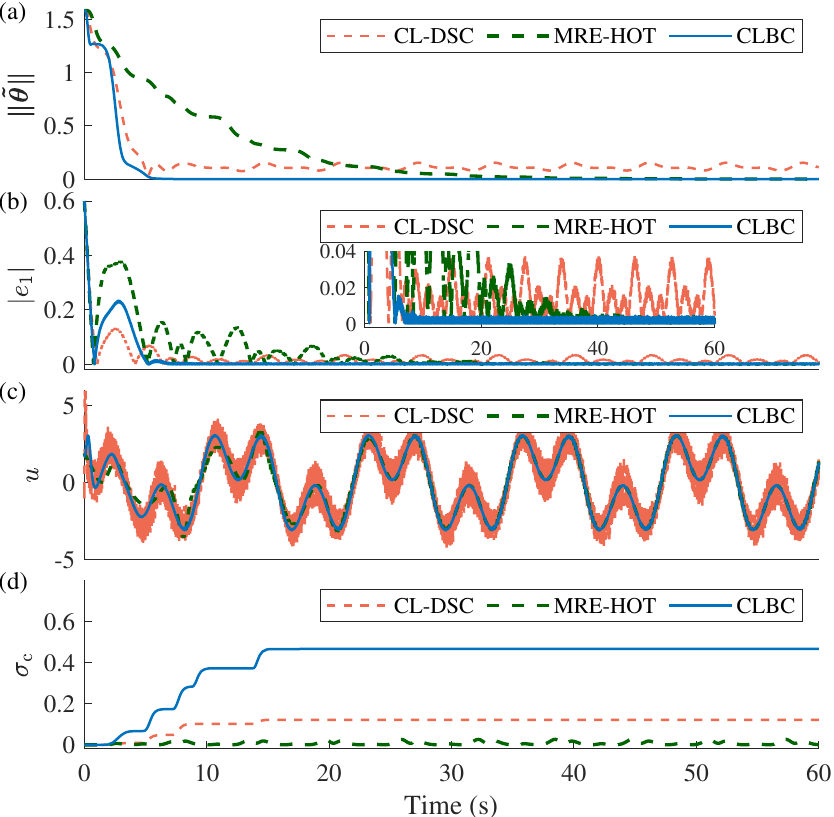}
\caption{Performance comparisons of three controllers for the tracking problem under the PE condition. (a) The estimation error norms $\|\tilde{\bm\theta}\|$. (b) The absolute tracking errors $|e_1|$. (c) The control inputs $u$. (d) The exciting strengths $\sigma_{\rm c}$.}
\label{fig:case1}
\end{figure}

\textit{Case 1: Tracking with PE.}
Consider a tracking problem under PE with the reference trajectory $y_{\rm r} = 1.5\sin(0.5t)$, the desired parameter $\bm\theta = [0.1, 0.5, 1.5]^T$, and the initial state $\bm x(0)$ $=$ $[0.6,0,0]^T$. 
Performance comparisons of the three controllers are depicted in Fig. \ref{fig:case1}. 
It is observed that the estimation error $\tilde{\bm\theta}$ by the CL-DSC has a steady-state error since the time derivative of each virtual control input is estimated by the filter $L(s)$ [see Fig. \ref{fig:case1}(a)], 
$\tilde{\bm\theta}$ by the MRE-HOT converges to $\bm 0$ after running 30 s [see Fig. \ref{fig:case1}(a)] as PE is fulfilled in this case, and $\tilde{\bm\theta}$ by the proposed CLBC exhibits the rapid convergence to $\bm 0$ [see Fig. \ref{fig:case1}(a)], which validates its strong learning capability. 
Besides, the proposed CLBC exhibits much better tracking performance than the CL-DSC and MRE-HOT [see Fig. \ref{fig:case1}(b)] because the exciting strength $\sigma_{\rm c}$ of the CLBC is monotonic non-decreasing and keeps a high level throughout [see Fig. \ref{fig:case1}(d)]. 
Moreover, the control inputs $u$ by the proposed CLBC and the MRE-HOT are not affected by the measurement noise throughout and are comparable after their estimation errors $\tilde{\bm\theta}$ converge to $\bm 0$ [see Fig. \ref{fig:case1}(c)], but $u$ by the CL-DSC is seriously polluted by the measurement noise [see Fig. \ref{fig:case1}(c)], which results in much worse tracking performance [see Fig. \ref{fig:case1}(b)].

\begin{figure}[!t]
\centering
\includegraphics[width=3.4in]{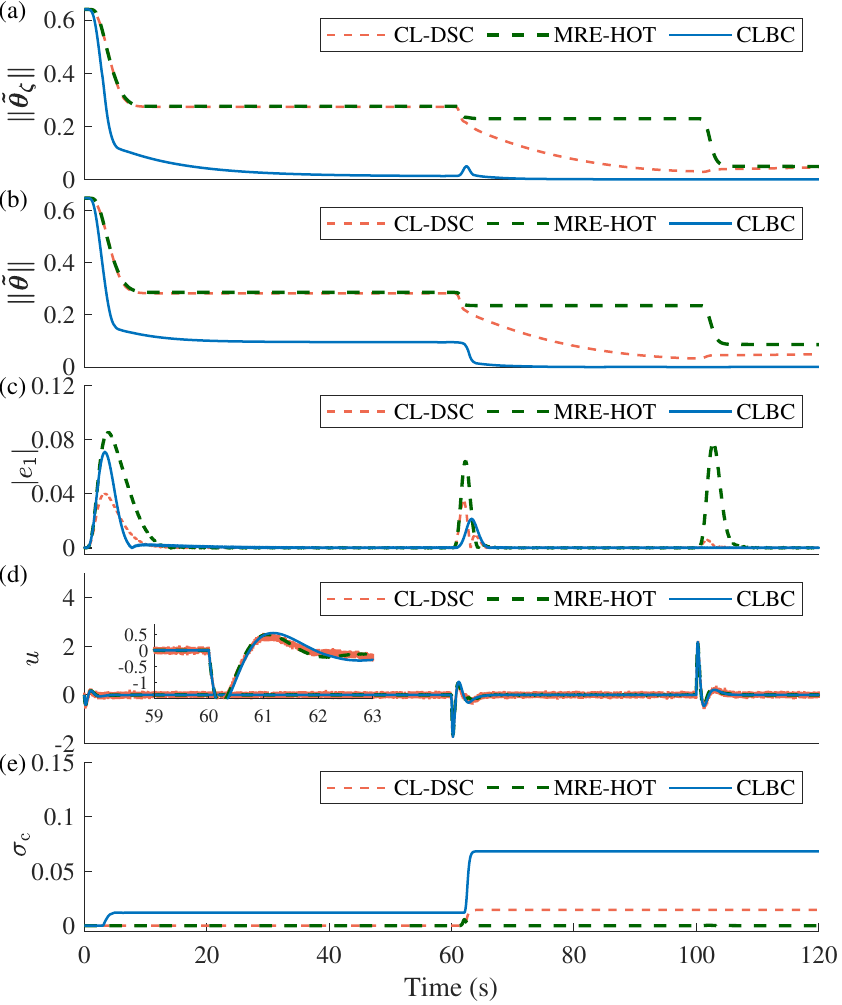}
\caption{Performance comparisons of three controllers for the regulation problem under partial IE or IE condition. (a) The partial estimation errors $\|\tilde{\bm\theta}_\zeta\|$. (b) The estimation errors $\|\tilde{\bm\theta}\|$. (c) The absolute tracking error norms $|e_1|$. (d) The control inputs $u$. (e) The exciting strengths $\sigma_{\rm c}$.}
\label{fig:case2}
\end{figure}


\textit{Case 2: Regulation with partial IE or IE.}
Consider a regulation problem under the partial IE or IE condition with the desired parameter $\bm\theta = [0.4, 0.5, 0.1]^T$ and the initial state $\bm x(0)=\bm 0$, where the reference trajectories $y_{\rm r}$, $\dot{y}_{\rm r}$, $\ddot{y}_{\rm r}$, and $\dddot{y}_{\rm r}$ are generated by a reference model $y_{\rm r}(t)$ $=$ $\frac{a_{0}}{a(s)}[r(t)]$ with $a_{0} = 16$, $a(s) = s^4+8s^3+24s^2+32s+16$, and 
\begin{align*}
    r(t) = \left\{\begin{array}{ll}
       -0.3,  &  t<60 \\
       -1.5,  &  60\leq t<100 \\
       0,     & t\geq 100 \\
    \end{array}\right.
\end{align*}
such that only the first two regressor channels are sufficiently excited during $t$ $\in$ [0,60) s, and IE is established after $t=$ 60 s. Performance comparisons of the three controllers are depicted in Fig. \ref{fig:case2} with $\tilde{\bm\theta}_\zeta$ $:=$ $[\tilde\theta_1, \tilde\theta_2]^T$. It is observed that the CL-DSC exhibits the rapid convergence of the estimation error $\tilde{\bm\theta}$ at $t\in[60, 120]$ s due to the establishment of IE but still has a steady-state error [see Fig. \ref{fig:case2}(b)], the MRE-HOT does not show the convergence of $\tilde{\bm\theta}$ to $\bm 0$ after 60 s [see Fig. \ref{fig:case2}(b)] due to the lack of PE, and  
the proposed CLBC shows the convergence of partial elements $\tilde\theta_1$ and $\tilde\theta_2$ to 0 at $t\in[0, 60)$ s and then all elements $\tilde\theta_i$ ($i=1$ to $3$) to 0 after $t = 60$ s [see Figs. \ref{fig:case2}(a) and (b)], which is consistent with the result in Theorem 1. 
Regarding the tracking performance, the proposed CLBC owns the highest tracking accuracy after $\tilde{\bm\theta}$ converges to $\bm 0$ [see Fig. \ref{fig:case2}(c)]. 
Moreover, the control inputs $u$ by the proposed CLBC and the MRE-HOT are comparable in this case, and the control input $u$ by the CL-DSC is sensitive to the measurement noise [see Fig. \ref{fig:case2}(d)]. 
It is worth noting that the exciting strengths $\sigma_{\rm c}$ by the CL-DSC and MRE-HOT are $0$ at $t\in[0,60)$ s due to the weakly excited channel $\bm\phi_3$ [see Fig. \ref{fig:case2}(e)], but their different control designs lead to obviously different tracking results as shown above.

\section{Conclusions}\label{Conclusions}

This paper has presented a CLBC with HOTs, which allows the exact implementation of the high-order time derivatives of parameter estimates, and enables the exponential stability of the closed-loop system with parameter convergence under the much weaker condition of IE or partial IE. Simulation studies have validated that the proposed CLBC greatly outperforms state-of-the-art backstepping controllers in both parameter estimation and control. Further work would focus on nonlinear parameterization and robot control based on the proposed method.


\bibliographystyle{IEEEtran}        
\balance
\bibliography{IEEEabrv,CLBC_ref}

\end{document}